\documentclass[aps,prl,reprint,superscriptaddress,amsmath,amssymb]{revtex4-1}
\usepackage{graphicx}% Include figure files
\usepackage{dcolumn}% Align table columns on decimal point
\usepackage{bm}% bold math
\usepackage{verbatim}

\begin{document}

%\begin{comment}
\title{Antiferromagnetic spin Seebeck effect}

\author{Stephen M. Wu}
	\email{swu@anl.gov}
\affiliation{%
Materials Science Division, Argonne National Laboratory, Argonne, Illinois 60439, USA
}%

\author{Wei Zhang}
\affiliation{%
Materials Science Division, Argonne National Laboratory, Argonne, Illinois 60439, USA
}%

\author{Amit KC}
\affiliation{%
Department of Physics and Astronomy, West Virginia University, Morgantown, West Virginia 26506, USA
}%

\author{Pavel Borisov}
\affiliation{%
Department of Physics and Astronomy, West Virginia University, Morgantown, West Virginia 26506, USA
}%

\author{John E. Pearson}%
\affiliation{%
Materials Science Division, Argonne National Laboratory, Argonne, Illinois 60439, USA
}%

\author{J. Samuel Jiang}
\affiliation{%
Materials Science Division, Argonne National Laboratory, Argonne, Illinois 60439, USA
}%

\author{David Lederman}
\affiliation{%
Department of Physics and Astronomy, West Virginia University, Morgantown, West Virginia 26506, USA
}%

\author{Axel Hoffmann}
\affiliation{%
Materials Science Division, Argonne National Laboratory, Argonne, Illinois 60439, USA
}%

\author{Anand Bhattacharya}%

\affiliation{%
Materials Science Division, Argonne National Laboratory, Argonne, Illinois 60439, USA
}%

\date{\today}% It is always \today, today,
             %  but any date may be explicitly specified

\begin{abstract}
We report on the observation of the spin Seebeck effect in antiferromagnetic MnF$_2$. A device scale on-chip heater is deposited on a bilayer of MnF$_2$ (110) (30 nm)/Pt (4 nm) grown by molecular beam epitaxy on a MgF$_2$ (110) substrate.  Using Pt as a spin detector layer it is possible to measure thermally generated spin current from MnF$_2$ through the inverse spin Hall effect. The low temperature (2 - 80 K) and high magnetic field (up to 140 kOe) regime is explored. A clear spin flop transition corresponding to the sudden rotation of antiferromagnetic spins out of the easy axis is observed in the spin Seebeck signal when large magnetic fields ($\textgreater$9 T) are applied parallel the easy axis of the MnF$_2$ thin film. When magnetic field is applied perpendicular to the easy axis, the spin flop transition is absent, as expected. 
\end{abstract}

\pacs{}% PACS, the Physics and Astronomy
                             % Classification Scheme.
%\keywords{Suggested keywords}%Use showkeys class option if keyword
                              %display desired
\maketitle 

The field of spin caloritronics has recently attracted a large amount of attention as a possible new direction for the world of spintronics \cite{bauer2012spin}. In spin caloritronic devices: information is transmitted by spin current instead of electrical current, the medium that carries spin current can be a magnetic insulator instead of an electrical conductor, and the primary driver of current is a thermal gradient instead of an electric field. The longitudinal spin Seebeck effect (SSE) lies at the center of this burgeoning field as the primary method of thermal spin current generation from magnetic insulators \cite{uchida2010observation,uchida2013longitudinal,ramos2013observation,
meier2013thermally,uchida2010longitudinal}.

Recently, it was discovered that in addition to ferromagnetic and ferrimagnetic insulators, it is also possible to generate spin current  through the SSE from insulating paramagnetic materials \cite{wu2015paramagnetic}. In these systems (Gd$_3$Ga$_5$O$_{12}$, DyScO$_3$) antiferromagnetic (AFM) interactions exist but fail to achieve long range ordering above a nominal AFM ordering temperature, and spin current generation is presumed to be due to short range interactions. This immediately leads to the question of whether thermal spin current generation is possible from the AFM phase itself. Spin current generation using insulating antiferromagnets alone has only been theoretically predicted \cite{ohnuma2013spin,cheng2014spin,brataas2015heat} without experimental observation until this work.

In this letter, we report on the thermal generation of spin current from the insulating AFM MnF$_2$ through the longitudinal spin Seebeck effect. This effect is due to thermal spin wave excitations from a material with a well defined spin wave spectrum, thus showing that in addition to ferromagnetic spin waves, antiferromagnetic spin waves can be used to generate spin current as well. Since AFM materials are free of stray fields, they are more immune to parasitic magnetic effects that may occur as spintronic device scaling becomes more important in future applications. AFM insulators are also far more common than the ferrimagnetic insulators typically used in spin Seebeck experiments, therefore opening a new larger class of materials for use in spin caloritronic devices.

\begin{figure}[b]
\includegraphics[width=3.4in,trim =0.35in 0.75in 0.35in 1.20in,clip=true]{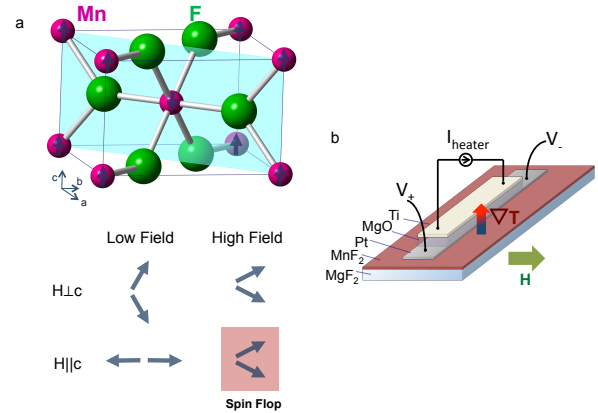}
\caption{\label{fig1} (a) The crystal structure of MnF$_2$ is presented with AFM spin structure overlaid on Mn$^{2+}$ ions. The (110) thin film crystal orientation plane is highlighted in blue. The spin flop transition in MnF$_2$ is presented. (b) Device schematic outlining a typical spin Seebeck device geometry. }
\end{figure}	 

MnF$_2$ has a tetragonal crystal structure, and an AFM Neel temperature of 67.7 K \cite{akutsu1981specific} with uniaxial anisotropy due to dipole interactions between Mn$^{2+}$ that causes a magnetic easy-axis along the c-axis direction as schematically shown in Fig. 1(a) \cite{keffer1952anisotropy,barak1978magnetic,nikotin1969magnon}. When a magnetic field is applied along the easy axis that exceeds a critical field H$_C$, the spins of both antiferromagnetic sublattices suddenly rotate and align mostly perpendicular to the c-axis in a canted state [Fig. 1(a)]. The detection of this abrupt spin flop transition in the SSE measurement is the primary evidence used to verify that a spin current is being generated by antiferromagnetic MnF$_2$. Once in the spin flopped state, the spins cant further in the direction of the magnetic field. This canted moment is $\sim$ 0.4 $\mu_B$/Mn at $\sim$90 kOe and is about 8\% of the sublattice magnetization \cite{felcher1996antiferromagnetic}.

The MnF$_2$ thin film, with an approximate thickness of 30 nm, was grown on a single crystal MgF$_2$ (110) substrate by molecular beam epitaxy (MBE). A 4 nm thick top Pt electrode film was prepared by sputtering ex-situ immediately after the deposition of MnF$_2$. The crystal structure of MnF$_2$ is shown schematically in Fig. 1(a) with the (110) plane highlighted. The surface of the film is nominally magnetically compensated, as seen in the schematic, but is likely more complicated in reality due to interfacial surface roughness. Details of the growth and characterization of the thin film are included in the supplementary information \cite{sup}.

Device structures were patterned using photolithography and argon ion milling	to etch the Pt layer into 300 $\mu$m x 10 $\mu$m bar structures oriented parallel and perpendicular to the c-axis. On top of this, a 100 nm electrical insulating layer of MgO and a 50 nm layer of electrically resistive Ti was deposited to serve as the heater for the device. A schematic of the device used in this experiment is presented in Fig. 1(b). This on-chip heating technique allows us to access lower temperatures (2 K) and higher magnetic fields (140 kOe) by easily integrating these devices into conventional superconducting magnet setups. 

\begin{figure}[ht]
\includegraphics[width=3.4in,trim =0in 0.9in 0.5in 0.5in,clip=true]{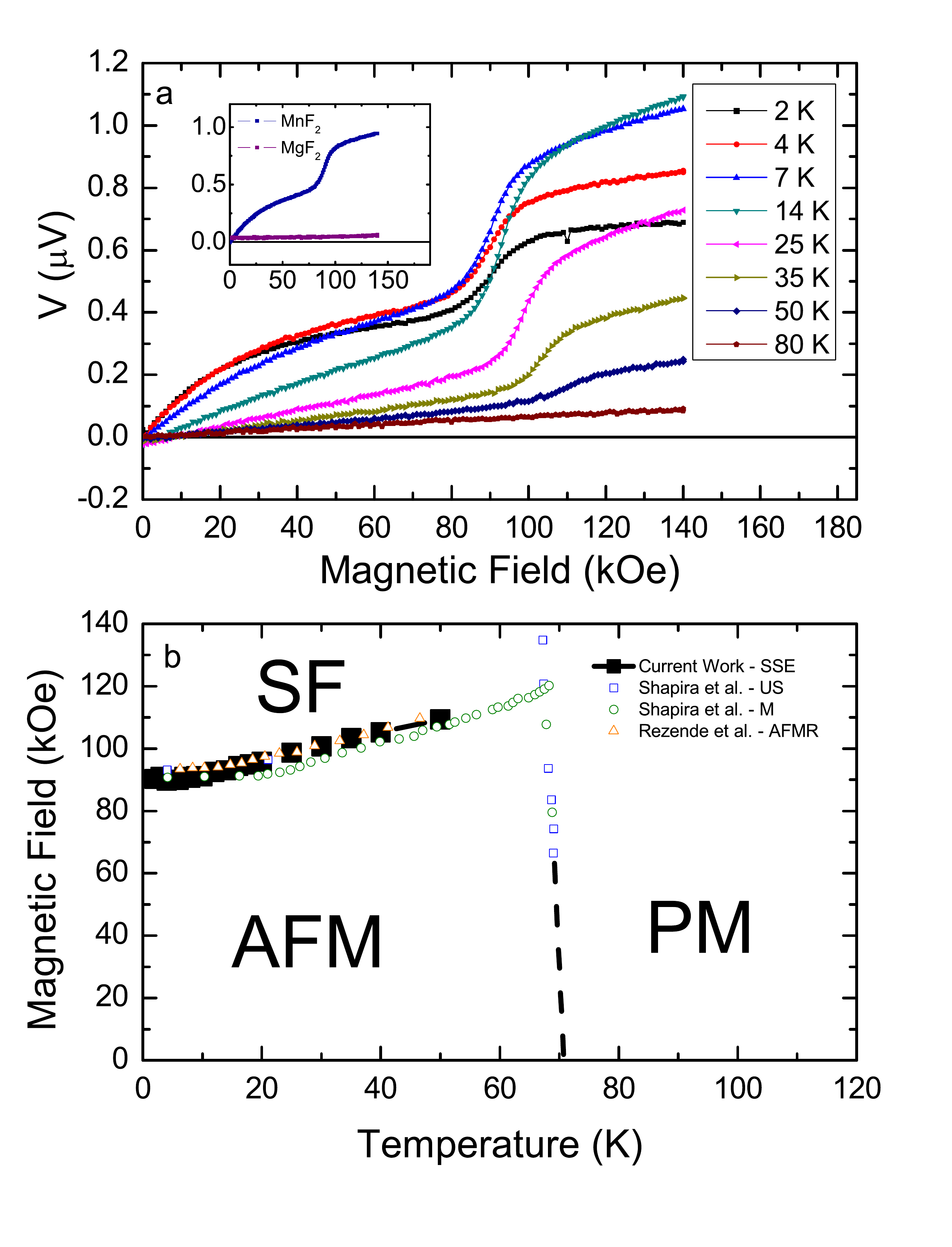}
\caption{\label{fig2} Spin Seebeck voltage response curves from MnF$_2$ are shown with magnetic field applied parallel to the c-axis in (a).  A control experiment is performed with a bare MgF$_2$ substrate with Cu (2 nm)/Pt (4 nm) under the same conditions at 5 K showing no measurable effect (inset). By mapping the spin flop transition from (a), a phase diagram for MnF$_2$ is reproduced in (b). This data is compared to data from Shapira et al. using ultrasonic attenuation (US), and differential magnetization (M), and Rezende et al. from antiferromagnetic resonance (AFMR) \cite{shapira1970magnetic, rezende1977stability}.    }
\end{figure}	

A constant voltage of 1 V$_{rms}$ was applied at 3 Hz to the $\sim$250 $\Omega$ heater layer over a 1000 $\Omega$ bias resistor while magnetic field was applied along the c-axis. In this measurement, to detect the spin current due to spin components along the c-axis,  the Pt bar was patterned perpendicular to the c-axis, which lies within the plane of the film. The resulting lock-in detected signal on the 90$^{\circ}$ out-of-phase channel at the 2nd harmonic (to isolate effects only due to heating) is presented in Fig. 2a for temperatures between 2 and 80 K. Here, the spin flop transition is clearly seen for temperatures below T$_N$, and qualitatively matches magnetization data for bulk MnF$_2$ except for a curvature that develops at low temperature \cite{jacobs1961spin}.  At 80 K, above T$_N$, only a linear voltage response is measured, likely due to the paramagnetic spin Seebeck effect \cite{wu2015paramagnetic} due to the size of the signal. Below T$_N$ there is a linear component to the voltage response until the spin flop transition H$_C$ where a large jump in signal is measured. As the temperature is lowered, a non-linear signal grows. This signal could be an intrinsic effect of the antiferromagnetic spin Seebeck effect due to the magnetic field induced splitting of the two antiferromagnetic magnon branches, which is well supported by recent theoretical work based on the magnon spin current theory of the SSE, where both the shape and temperature dependence of this effect is reproduced \cite{rezendeAFMSSEtheory}. As a control, the same measurement is performed on a bare MgF$_2$ substrate, which resulted in no response, and thus eliminates the possibility that this is a spurious effect from a paramagnetic substrate [inset in Fig. 2(a)].  The measured phase diagram is shown in Fig. 2(b) and compared with historical data on the spin flop boundary using multiple different techniques on bulk single crystal samples \cite{shapira1970magnetic,rezende1977stability}. The degree to which our spin Seebeck  measurements agree with bulk single crystal data suggests that our thin film samples are of high quality, and that there is a low likelihood that the measured effect is due to proximity magnetism induced into the Pt layer since the magnetic properties of MnF$_2$ are unlikely to be transferred one-to-one. To completely eliminate this possibility, control measurements were performed on another MnF$_2$ thin film sample using either W (4 nm) or Cu (2.5 nm)/Pt (4 nm) as the spin detector layer. The resulting SSE signal under the same heating conditions shows the same spin flop behavior \cite{sup}.

\begin{figure}[ht]
\includegraphics[width=3.4in,trim =0in 0in 0.5in 0.5in,clip=true]{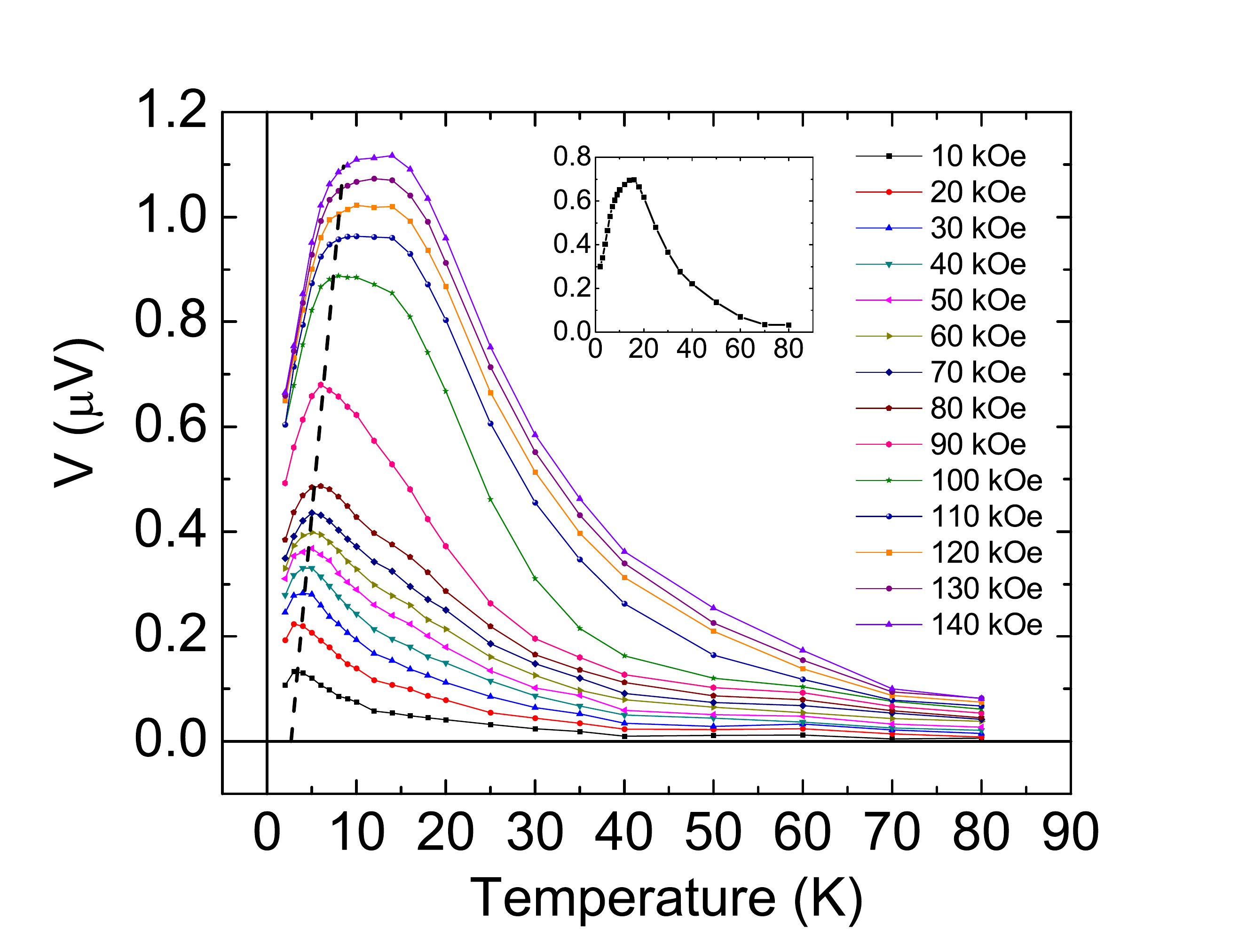}
\caption{\label{fig3} Temperature dependence of the spin Seebeck voltage response at various magnetic fields.  The inset represents the 140 kOe data with the 70 kOe subtracted out to judge the temperature dependence of just the spin-flopped phase.   }
\end{figure}	 

The temperature dependence of the measurements taken in Fig. 2(a) are presented in Fig. 3 for several different magnetic fields. Since the resistance of the Ti heater layer changes with temperature, the power applied to the heater changes approximately 10.6\% from 80 K to 2 K (0.177 mW$_{rms}$-0.160 mW$_{rms}$), but this effect is much smaller than the magnitude in change of the voltage signal \cite{sup}.  The temperature dependence shows a peak at low temperatures at all magnetic fields. At low magnetic fields there appears to be a low temperature peak whose position increases in temperature with magnetic fields strength. At fields above the spin flop transition this peak becomes broader and approximately matches the peak in thermal conductivity of MnF$_2$ from literature \cite{slack1961thermal}. Many longitudinal spin Seebeck systems have a correlation between the size of the spin Seebeck signal and the thermal conductivity, which is believed to be a consequence of magnon-phonon interaction \cite{uchida2012thermal,rezende2014magnon}. In our device geometry there is typically an inverse dependence on the size of the signal to the thermal conductivity of the thin film since a constant power is applied to the material instead of a constant temperature difference $\Delta$T \cite{wu2014unambiguous,wu2015spin,wu2015paramagnetic}. Here, V$\propto \Delta$T$\propto \frac{P}{\kappa}$, where V is the measured voltage due to the inverse spin Hall effect, $P$ is the applied power, and $\kappa$ is the thermal conductivity of the film. Since our measurement suggests V scales with $\kappa$, there is minimally a stronger than linear dependence of the spin Seebeck signal size on $\kappa$.  This could be due to especially weak interaction between magnons and phonons in this system due to higher frequency gapped AFM magnons, leading to a larger temperature difference between non-equilibrium  phonon and magnon populations \cite{sanders1977effect}. Both the heat capacity and thermal conductivity of MnF$_2$ is dominated by phonon conduction, and therefore the effect of magnetic field on the thermal conductivity of MnF$_2$ is negligible and cannot account for the spin flop behavior in the SSE \cite{sanders1977thermal,sanders1977effect}. The lack of magnon thermal conduction is also evidence that the magnon-phonon relaxation times are long due to weak interaction in MnF$_2$ \cite{sanders1977thermal}. The inset of Fig 3. shows the data from 120 kOe with the contribution at 70 kOe subtracted to isolate the temperature dependence of  the SSE in the spin-flopped phase. The data show a sharper peak at $\sim$20 K, suggesting that the SSE in the spin-flopped phase is strongly correlated with the MnF$_2$ thermal conductivity.

To confirm that the origin of the jump in the spin Seebeck signal is from the spin flop transition, measurements were made on a separate device fabricated simultaneously on the same film, with the pattern oriented 90$^{\circ}$ to the original device. In this device, spin current due to spin components perpendicular to the c-axis are detected. The voltage response from the new device, performed under the same conditions as in Fig. 2(a), is compared to the data for magnetic field parallel to the c-axis. The results are summarized in Fig. 4, where the jump in spin Seebeck signal is absent with magnetic field in the $\perp$ to c-axis direction, while still present in the $\parallel$ to c-axis case. At 80 K, above T$_N$, both signals are roughly equivalent. As the temperature is lowered below $T_N$, the signal in the $\parallel$c device is lower than in the $\perp$c device for $H<H_C$, but the two signals roughly agree with each other for $H>H_C$.  Because the two devices are identical except for the direction of the Pt bar, it is unlikely that the observed phenomena are due to proximity magnetic interactions or diffusion of magnetic ions into the Pt layer since this anisotropic behavior is specific to only MnF$_2$. 
	
\begin{figure}[ht]
\includegraphics[width=3.4in,trim =0in 0.3in 0.5in 3.5in,clip=true]{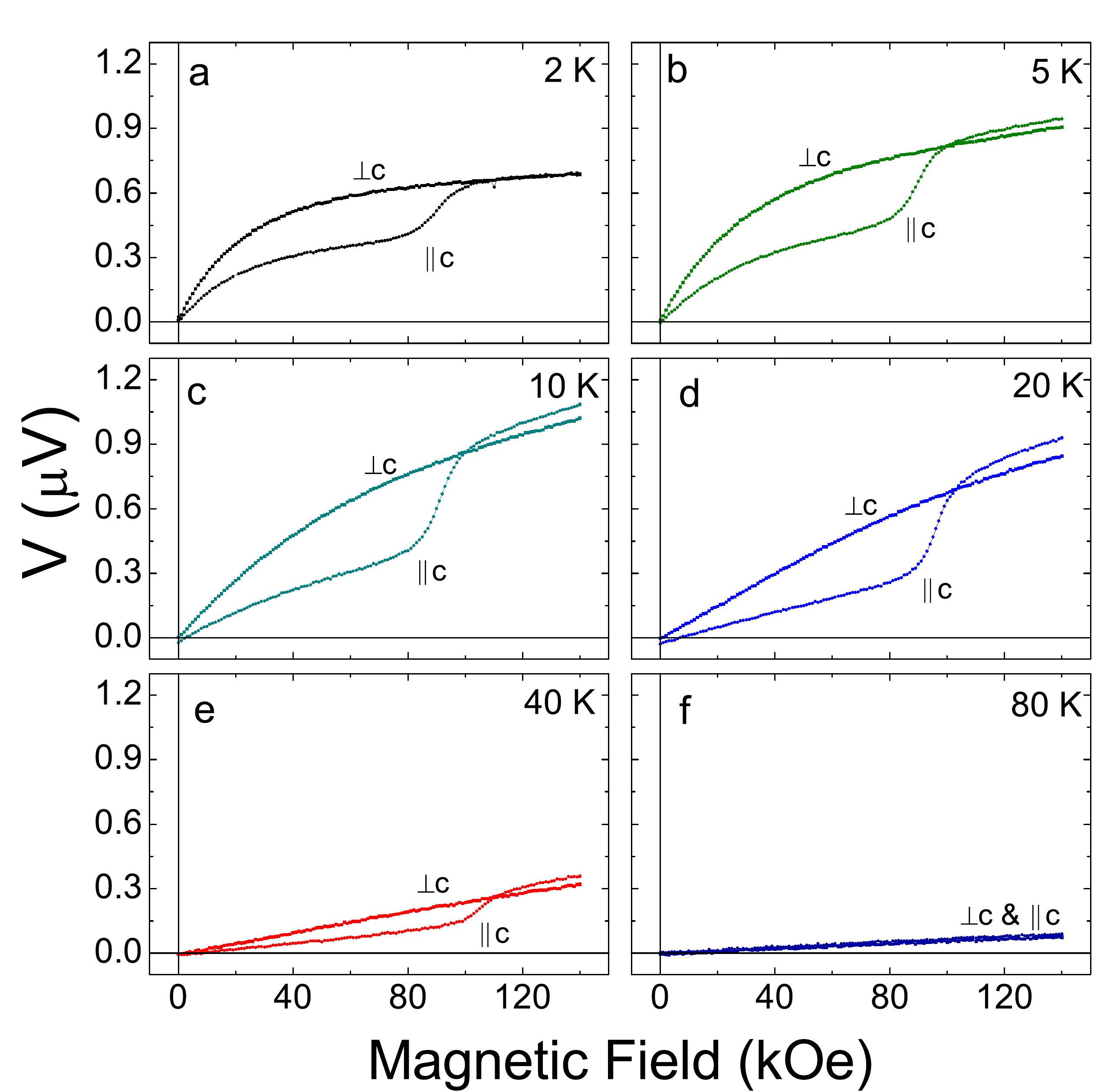}
\caption{\label{fig4} (a)-(f) Spin Seebeck voltage responses on two devices, one aligned to detect spin current parallel to the c-axis, and one aligned to detect spin current perpendicular to the c-axis. The spin flop transition is only present in the parallel configuration. }
\end{figure}	 

Current theories on the origin of the spin Seebeck effect involve a non-equilibrium population of magnons accumulating at the interface between the magnetic insulator and metallic spin detector layer \cite{xiao2010theory,adachi2013theory,rezende2014magnon}. This could be due to several mechanisms, including bulk magnon diffusion \cite{rezende2014magnon}, or building a steady state non-equilibrium magnon population due to a finite magnon-phonon relaxation time \cite{schreier2013magnon}.This non-equilibrium population of magnons interacts with  electrons in the spin detector layer through an incoherent thermally excited spin pumping process where spin current is transferred to the metallic spin detector layer. Since this is a thermally generated effect, the population of excited magnons depends strongly on the entire magnon spectrum. In AFM systems, the same SSE mechanism can occur with antiferromagnetic spin waves, which have different spectra when compared to ferromagnets. 

In MnF$_2$, the spin wave spectrum has a gap of 1.081 meV, measured through neutron scattering \cite{low1964measurement,nikotin1969magnon} and antiferromagnetic resonance experiments \cite{johnson1959antiferromagnetic}. It is known that the two degenerate bands from the individual Mn sublattices split in a magnetic field applied along the easy axis \cite{johnson1959antiferromagnetic}. The behavior of the k=0 spin wave mode under magnetic field can be obtained from antiferromagnetic resonance experiments, and provides a guide to the behavior of the rest of the spectrum since inelastic neutron scattering data at high magnetic fields are absent \cite{hagiwara1999complete}. It is possible to identify spin current generation from a system like MnF$_2$ since there is a large abrupt change in the spin wave spectrum through the spin flop transition, which can be inferred from antiferromagnetic resonance experiments \cite{ross2015antiferromagentic}. Both theoretical and experimental evidence for changes in the spin Seebeck effect due to changes in magnon branch degeneracy have been reported for compensated ferrimagnetic systems \cite{geprags2014origin,ohnuma2013spin}. This type of change in the MnF$_2$ spin flop transition could lead to a change in the net spin current and an abrupt change in the voltage response like the one observed in our devices. The magnitude of the signal measured is larger in size to equivalent measurements in ferromagnetic materials \cite{wu2014unambiguous,wu2015spin,sup}, which suggests that the origin of this effect is not solely due to the AFM canting which only amounts to 8\% of the individual sublattice magnetizations after the spin flop. Regardless, since the dominant exchange interaction in both the AFM and spin flop phases are antiferromagnetic, thermal spin current generation in this system is likely mediated by antiferromagentic spin waves. 

Independent concurrent work by Seki et al. on the antiferromagnet Cr$_2$O$_3$ shows a similar spin flop transition in spin Seebeck signal on large bulk single crystals \cite{seki2015thermal}. Some differences  between the results by Seki et al. and the results of this work are that at even the lowest temperatures there is a signal from MnF$_2$ below the spin flop transition that is absent in Cr$_2$O$_3$. Additionally, the SSE in MnF$_2$ is larger with an estimated spin Seebeck coefficient \cite{sola2015evaluation} of 4.5 $\mu$V/K at 35 K and 14 T when literature values for thermal conductivity are used, whereas in Cr$_2$O$_3$ this value is closer to 0.015 $\mu$V/K for the same temperature and field. This may be due to the effect of the differences in the intrinsic magnonic properties of the two materials (AFM exchange, magnon-phonon relaxation, magnon dispersion), or the larger canted moment at the spin flop transition in MnF$_2$ ($\sim$ 0.4 $\mu_B$/Mn) vs. Cr$_2$O$_3$ ($\sim$ 0.02 $\mu_B$/Cr). At 15 K and 14 T, using the same assumption that the thin film thermal conductivity is equivalent to the literature values on bulk single crystal samples \cite{slack1961thermalmnf2}, the spin Seebeck coefficient is calculated to be 41.2 $\mu$V/K, which is larger than even the largest values of the SSE in bulk single crystal YIG at low temperatures (4.6 $\mu$V/K) \cite{criticalsupression}. One caveat in this type of analysis is that the thin film cross plane thermal conductivity may be smaller than the bulk value and is challenging to quantify directly without specialized techniques. However, this large response is supported by our own comparison measurements of the SSE on thin film YIG, where after considering geometric factors, the voltage generated by MnF$_2$ is 50 times larger per unit power \cite{sup}.

In conclusion, we have shown that in thin film antiferromagnetic MnF$_2$ it is possible to measure the spin flop transition when magnetic field is applied along the magnetic easy axis using the spin Seebeck effect. This is direct evidence that the voltage measured from the spin detector layer is a direct consequence of spin current generation from the antiferromagnetic material. The source of this spin current generation could be due to magnetic field induced changes in the spin wave spectra of the material. Further work on different antiferromagnetic systems with different spin wave and thermal properties could lead to more insights on the mechanism of non-equilibrium magnon generation, and provide a new class of materials to engineer into thermal spintronic device applications.

\begin{acknowledgments}
All work at Argonne was supported by the U.S. Department of Energy, Office of Science, Basic Energy Sciences, Materials Sciences and Engineering Division. The use of facilities at the Center for Nanoscale Materials, an Office of Science user facility, was supported by the U.S. Department of Energy, Basic Energy Sciences under contract No. DE-AC02-06CH11357. The work at WVU was supported by a Research Challenge Grant from the West Virginia Higher Education Policy Commission (HEPC.DSR.12.29), a grant from the National Science Foundation (grant No. DMR-1434897), and the WVU Shared Research Facilities. We thank S. M. Rezende for bringing the historical spin flop boundary information to our attention, and sharing his unpublished work with us.
\end{acknowledgments}

\end{document}